\documentclass[aps,prl,notitlepage,reprint,superscriptaddress,floatfix]{revtex4-1}

\usepackage{graphicx}
\usepackage{longtable}
\LTcapwidth=\textwidth
\usepackage{siunitx}
\usepackage{color}
\usepackage[colorlinks,citecolor=blue,urlcolor=blue,linkcolor=red]{hyperref}
\usepackage{lipsum}  
\usepackage{amsmath}
\usepackage{amsfonts}
\usepackage{bbm}
\usepackage{mathptmx}
\usepackage[utf8]{inputenc}
\usepackage{gensymb}

\begin{document}

\title{Reliable and Practical Computational Prediction of Molecular Crystal Polymorphs}

\author{Johannes Hoja}
\affiliation{Physics and Materials Science Research Unit, University of Luxembourg, L-1511 Luxembourg, Luxembourg}

\author{Hsin-Yu Ko}
\affiliation{Department of Chemistry, Princeton University, Princeton, NJ 08544, USA}

\author{Marcus A. Neumann}
\affiliation{Avant-garde Materials Simulation, 79100 Freiburg, Germany}

\author{Roberto Car}
\affiliation{Department of Chemistry, Princeton University, Princeton, NJ 08544, USA}

\author{Robert A. DiStasio Jr.}
\affiliation{Department of Chemistry and Chemical Biology, Cornell University, Ithaca, NY 14853, USA}

\author{Alexandre Tkatchenko}
\email[E-mail: ]{alexandre.tkatchenko@uni.lu}
\affiliation{Physics and Materials Science Research Unit, University of Luxembourg, L-1511 Luxembourg, Luxembourg}

\maketitle

\textbf{The ability to reliably predict the structures and stabilities of a molecular crystal and its (often numerous) polymorphs without any prior experimental information would be an invaluable tool for a number of fields, with specific and immediate applications in the design and formulation of pharmaceuticals~\cite{Cruz2015}. In this case, detailed knowledge of the polymorphic energy landscape for an active pharmaceutical ingredient yields profound insight regarding the existence and likelihood of late-appearing polymorphs~\cite{Bucar2015}, and hence has significant public health and economic implications. However, the computational prediction of the structures and stabilities of molecular crystal polymorphs is particularly challenging due to the high dimensionality of conformational and crystallographic space accompanied by the need for relative (free) energies to within $\approx 1$~kJ/mol per molecule. In this work, we combine the most successful crystal structure sampling strategy with the most accurate energy ranking strategy of the latest blind test of organic crystal structure prediction (CSP), organized by the Cambridge Crystallographic Data Centre (CCDC)~\cite{Reilly2016, Gibney2015}. Our final energy ranking is based on first-principles density functional theory (DFT) calculations that include three key physical contributions: (\textit{i}) a sophisticated treatment of Pauli exchange-repulsion and electron correlation effects with hybrid functionals, (\textit{ii}) inclusion of many-body van der Waals dispersion interactions, and (\textit{iii}) account of harmonic (and sometimes anharmonic) vibrational free energies. In doing so, this combined approach has an optimal success rate in producing the crystal structures corresponding to the five blind-test molecules and even predicts stable polymorphs that have not been observed to date. With this practical approach, we demonstrate the feasibility of obtaining reliable  structures and stabilities for molecular crystals of pharmaceutical importance, paving the way towards an enhanced fundamental understanding of polymorphic energy landscapes and routine industrial application of molecular CSP methods.}

Accurate and reliable CSP methods are able to furnish detailed knowledge of the energetic landscape corresponding to a given molecular crystal and its (often numerous) thermodynamically relevant polymorphs. With access to the structures and relative thermodynamical stabilities of such varied crystal-packing motifs, one can gain crucial insight into whether or not the existing structure of a pharmaceutical drug candidate is indeed the most thermodynamically stable solid form at ambient conditions. This in turn enables an informed and critical assessment of the potential risk associated with the assumed stable form disappearing at some point during the manufacturing process or consumable shelf life~\cite{Bucar2015}. In this case, a polymorph with similar stability but different (and often unwanted) properties could emerge as the dominant solid form---an event which can trigger a cascade of deleterious health-related, social, and financial repercussions. As such, the utilization of accurate and reliable computational CSP methods in conjunction with experimental polymorph screening efforts offers a comprehensive and sustainable solution to this grand challenge~\cite{Price2014}.

\begin{figure}[b!]
\includegraphics[width=0.85\columnwidth]{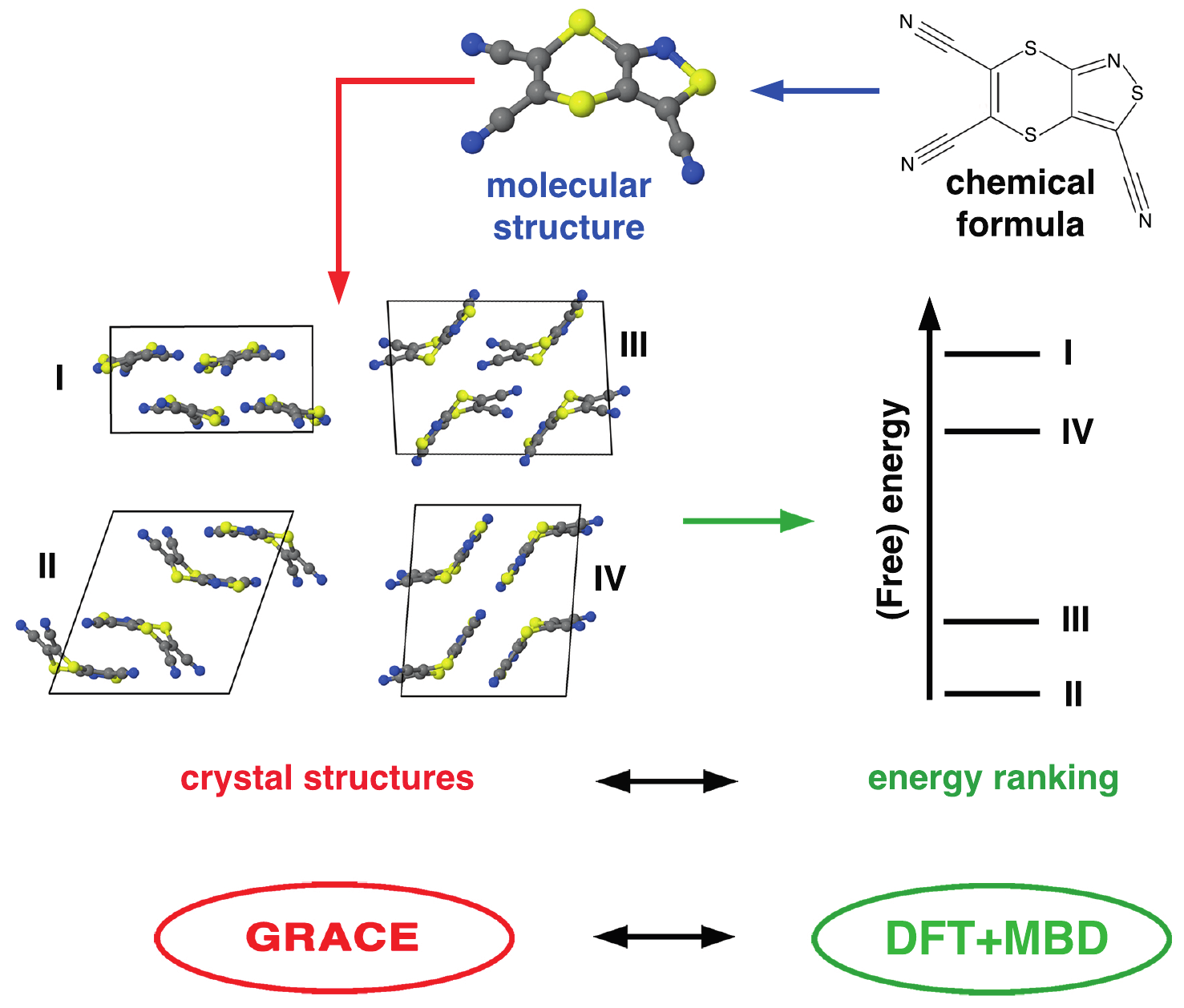}
\caption{\label{Fig:CSP} General overview of the crystal structure prediction (CSP) protocol. Starting with the 2D chemical formula for each molecule, this procedure generates molecular crystal structures and (free) energy rankings for all thermodynamically relevant polymorphs. In this work, we combine the crystal structure sampling strategy provided by the \textsc{Grace} software package with the highly accurate (free) energy ranking strategy provided by the first-principles based DFT+MBD framework.}
\end{figure}

In general, the accuracy and reliability of a given CSP methodology depends on two distinct but equally important theoretical aspects: (\textit{i}) sufficiently complete sampling of the conformational and crystallographic space spanned by a given molecular crystal and (\textit{ii}) sufficiently accurate ranking of the numerous low-energy polymorphs according to their relative thermodynamic stabilities~\cite{Reilly2016,Price2014}. In this regard, major advances have been made along both of these thrusts over the past few years, resulting in substantial progress in the field of molecular CSP~\cite{Beran2016,Yang2014,Nyman2015,Nyman2016,Schneider2016,Brandenburg2016,Price2017,Shtukenberg2017}. By and large, the most important benchmarks for assessing the utility of a given molecular CSP approach are the regular blind tests organized by the CCDC~\cite{Day2009,Bardwell2011,Reilly2016}, wherein participants predict the structure of a given molecular crystal based solely on the two-dimensional (2D) chemical formula for the individual molecule(s) involved. Over the past few decades, the chemical diversity and complexity of the CCDC blind test has gradually increased and now includes small and rigid molecules as well as elaborate polymorphic systems involving large and flexible molecules, salts, and co-crystals. A general overview of the protocol employed in a typical molecular CSP methodology is illustrated in Fig.~\ref{Fig:CSP}. Starting with the 2D chemical formula for each molecule, 3D molecular structures are first computed using standard geometry optimization techniques that are supplemented with additional sampling of all energetically relevant conformational isomers for flexible molecules. Next, a vast number of possible crystal-packing arrangements is generated by comprehensively sampling different intermolecular orientations, space groups, unit cell sizes, and molecular conformations. Finally, the generated crystal structures are ranked according to their relative (free) energies.

In this work, we demonstrate that an accurate, reliable, and computationally feasible protocol for the prediction of molecular crystal polymorphs can be obtained by combining the most successful crystal structure sampling strategy (Neumann \emph{et al.}) with the most successful first-principles energy ranking strategy (Tkatchenko \emph{et al.}) from the latest CCDC blind test~\cite{Reilly2016,Gibney2015}. In this regard, the approach for generating crystal structures by Neumann \emph{et al.} was able to correctly predict all experimentally observed structures (except for one) within the top 100 most stable structures, building on top of their major successes in previous blind tests~\cite{Day2009,Bardwell2011,Neumann2008}. The fact that several experimental structures could only be found with this approach again highlights the complexity associated with sufficiently sampling wide swaths of crystallographic space. In this approach, initial molecular crystal structures are created with a Monte Carlo parallel tempering algorithm that employs a tailor-made force field within the \textsc{Grace} software package. Following this initial screening, a set of candidate crystal structures are then reoptimized in a hierarchical and statistically-controlled process using dispersion-inclusive DFT~\cite{Neumann2005,Neumann2008,Neumann2008a,Reilly2016}. Beyond the robust sampling of the essential regions of crystallographic space, these initial energy rankings can be substantially improved upon by employing state-of-the-art first-principles methodologies as detailed below.

\begin{figure*}[t!]
\includegraphics[width=\textwidth]{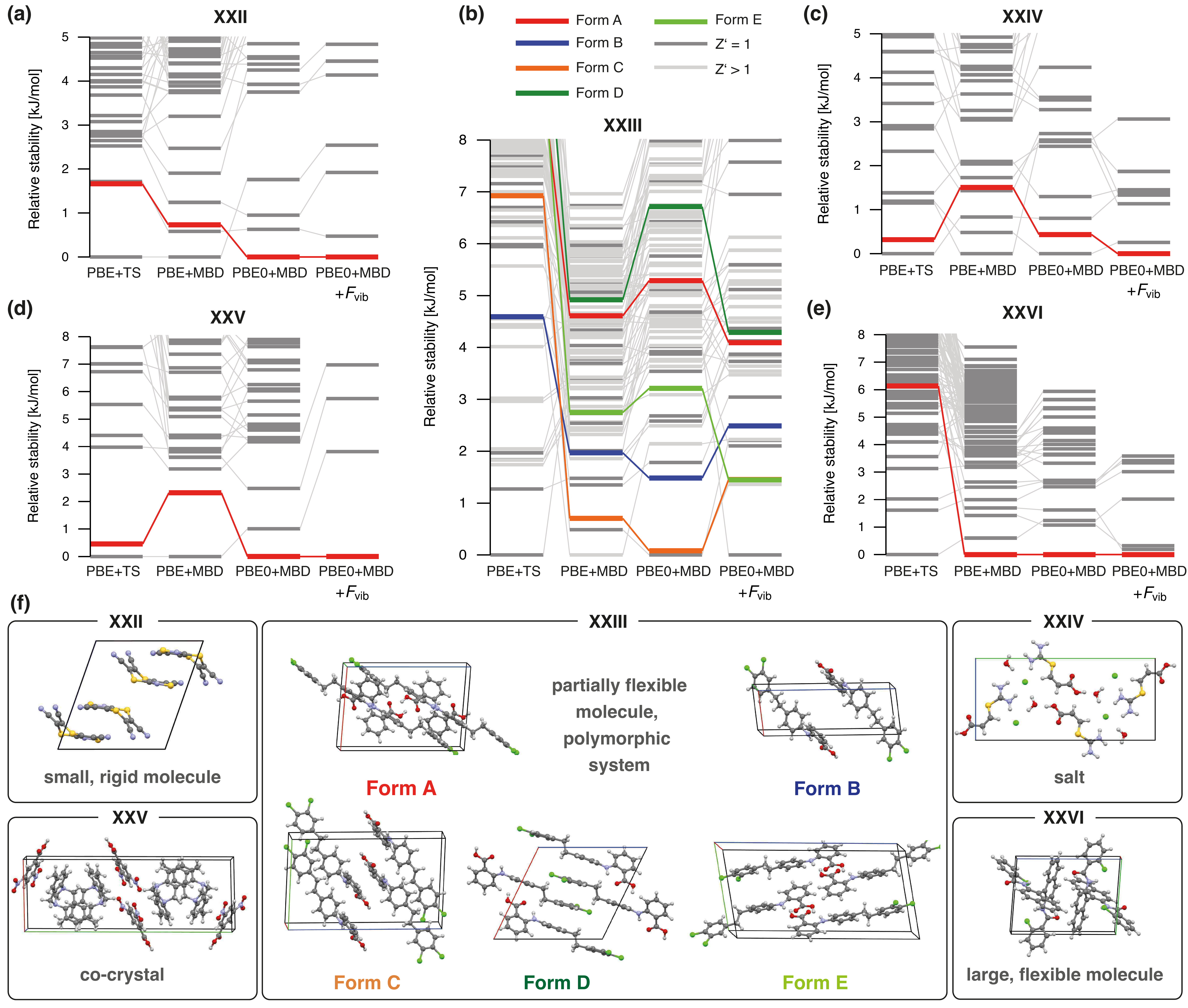}
\caption{\label{fig:ranking} Results from all steps in the present CSP stability ranking procedure for systems: \textbf{(a)} XXII, \textbf{(b)} XXIII, \textbf{(c)} XXIV, \textbf{(d)} XXV, and \textbf{(e)} XXVI. For each ranking, the energy of the most stable crystal structure defines the zero-of-the-energy. Experimentally observed structures are highlighted in color while all other structures are in gray. The final ranking for each system corresponds to the Helmholtz free energies at the PBE0+MBD+$F_{\rm vib}$ level, computed at the corresponding experimental temperatures: $150$~K for XXII, $240$~K for XXIV, and $300$~K for XXIII, XXV, and XXVI. All relative energies are reported per chemical unit, \textit{i.e.}, for XXII, XXIII, and XXVI, the energies are normalized per molecule, for XXIV and XXV, the energies are given per trimer and dimer, respectively. \textbf{(f)} Unit cells for all highlighted structures.}
\end{figure*}

To demonstrate this procedure, we start with the top 100 initial molecular crystal structures (for every system in the blind test) provided by Neumann \emph{et al.} (see Supplementary Information of Ref. \citenum{Reilly2016}). Form E of system XXIII is the only experimental structure that was not present in this set of initial structures and is included for completeness. We note in passing that this form was in fact generated by Neumann \emph{et al.}, but was located just outside the energetic window considered for the $Z'=2$ structures. In total, this set includes $501$ structures (with unit cell sizes ranging from $15$ to $992$ atoms) and therefore provides a large-scale benchmark structural database under realistic CSP conditions.

Based on these initial molecular crystal structures, we have developed a robust hierarchical first-principles approach for energetically ranking all relevant polymorphs. This approach is directly applicable to pharmaceutically relevant systems and includes three important theoretical aspects that are commonly neglected in typical CSP protocols: (\textit{i}) a sophisticated treatment of Pauli exchange-repulsion and electron correlation effects with hybrid functionals, (\textit{ii}) inclusion of many-body dispersion interactions and dielectric screening effects, and (\textit{iii}) an account of harmonic (and sometimes anharmonic) vibrational contributions to the free energy. In this regard, the hybrid PBE0 functional~\cite{Adamo1999} in conjunction with the many-body dispersion (MBD) model~\cite{Tkatchenko2012,DiStasio2012,Ambrosetti2014,Distasio2014,BloodForsythe2016} is able to predict absolute experimental lattice energies to within $1$~kcal/mol~\cite{Reilly2013,Roza2012} and relative stabilities of several polymorphic systems to within $1$~kJ/mol~\cite{Marom2013,Reilly2013,Reilly2015,Shtukenberg2017}. Hence, the PBE0+MBD approach is used for all calculations of static lattice energies. Geometry and lattice optimizations, as well as vibrational free energies are computed with the PBE functional~\cite{Perdew1996} in conjunction with the effective-pairwise Tkatchenko-Scheffler (TS) dispersion correction~\cite{Tkatchenko2009} (denoted as PBE+TS). A detailed description of the computational approaches employed in this work is available below in the \textit{Methods} section.

The stability rankings obtained for the five blind-test systems are shown in Fig.~\ref{fig:ranking} (with all structures and energies (Tables S6-S10) available in the \textit{Supplementary Information} (SI)). In this figure, we not only show the final stability rankings, but also several intermediary steps, in which one or more of the three aforementioned theoretical contributions are not accounted for in the rankings. The first ranking considers only static lattice energies computed at the PBE+TS level, while the second ranking accounts for beyond-pairwise many-body dispersion interactions (PBE+MBD). In the third ranking, we include a more sophisticated treatment of Pauli exchange-repulsion \textit{via} PBE0+MBD. In doing so, the deleterious effects of self-interaction error (a DFT artifact in which an electron interacts with itself) are significantly ameliorated, which leads to a substantial improvement in the description of electrostatic and charge-transfer effects. In the final ranking, we supplement the PBE0+MBD energies with harmonic vibrational free energy contributions at the PBE+TS level ($+F_{\rm vib}$). This leads to a stability ranking based on Helmholtz free energies which accounts for thermal entropic effects.

We first concentrate our discussion on systems XXII, XXIV, XXV, and XXVI. For all of these systems, our final stability ranking at the PBE0+MBD+$F_{\rm vib}$ level predicts the experimental structure as the most stable form---the ideal outcome of any CSP protocol. As seen from the intermediate stability rankings, all of the three previously mentioned theoretical effects are required to obtain this result. For example, Pauli exchange-repulsion (through the PBE0 functional) plays a crucial role for system XXII~\cite{Curtis2016}, while many-body dispersion effects are the most important factor for system XXVI. In addition, all structures with free energies that are within $1$~kJ/mol of the experimental structure are essentially minor variations of the latter (see SI), which demonstrates the robustness of our CSP approach in dealing with pharmaceutically-relevant systems like salts, co-crystals, and molecular crystals involving large and flexible molecules.

Now we focus our discussion on the most challenging system in the blind test (XXIII). This system involves a conformationally flexible molecule and has five experimentally confirmed polymorphs~\cite{Reilly2016}. The fact that this compound is also a former drug candidate~\cite{Simons2009} makes it an ideal testing ground for CSP of pharmaceutically-relevant molecules. The low-energy barriers between different conformers in this molecule lead to a fairly complex polymorphic landscape with numerous crystal structures located within a very small energy window. As shown in Fig.~\ref{fig:ranking}, the PBE+TS method is again insufficient for quantitative energy ranking predictions and places all experimentally observed structures within the top 11 kJ/mol---an energy window containing 84 structures. Each refinement of the energetic rankings changes their relative stabilities, with all experimental structures observed within the top $4.3$~kJ/mol ($\approx 1$~kcal/mol) in the final ranking with PBE0+MBD+$F_{\rm vib}$. At this level, all experimental structures were found within an energy interval of $3$~kJ/mol, the expected energy range associated with co-existing polymorphs~\cite{Cruz2015}. We note here that our procedure finds one structure (Str. N70) that is $\approx 1.5$~kJ/mol more stable than all experimentally observed structures, a remarkable finding that is discussed in more detail below.

\begin{figure}[t!]
\includegraphics[width=\columnwidth]{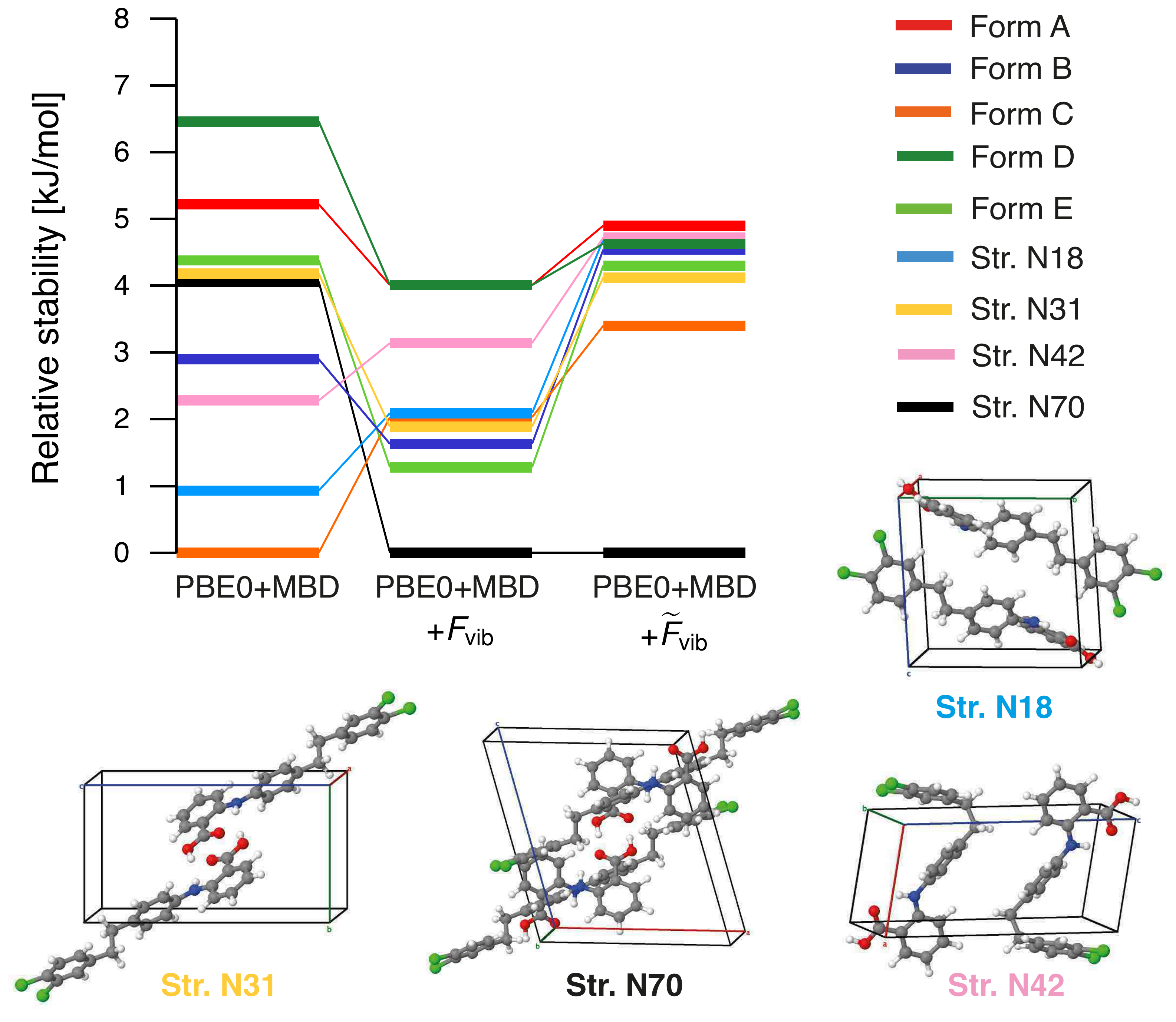}
\caption{\label{fig:csp2} Energetic rankings for all experimentally observed (Form A, B, C, D, E) and theoretically-predicted (N18, N31, N42, N70) structures for system XXIII. All energies were evaluated using thermally-expanded PBE+TS structures optimized at $300$~K with the quasi-harmonic approximation (QHA). The last two rankings include harmonic ($F_{\rm vib}$) and Morse anharmonic ($\widetilde{F}_{\rm vib}$) vibrational free energy contributions.}
\end{figure}

\begin{figure*}
\includegraphics[width=\textwidth]{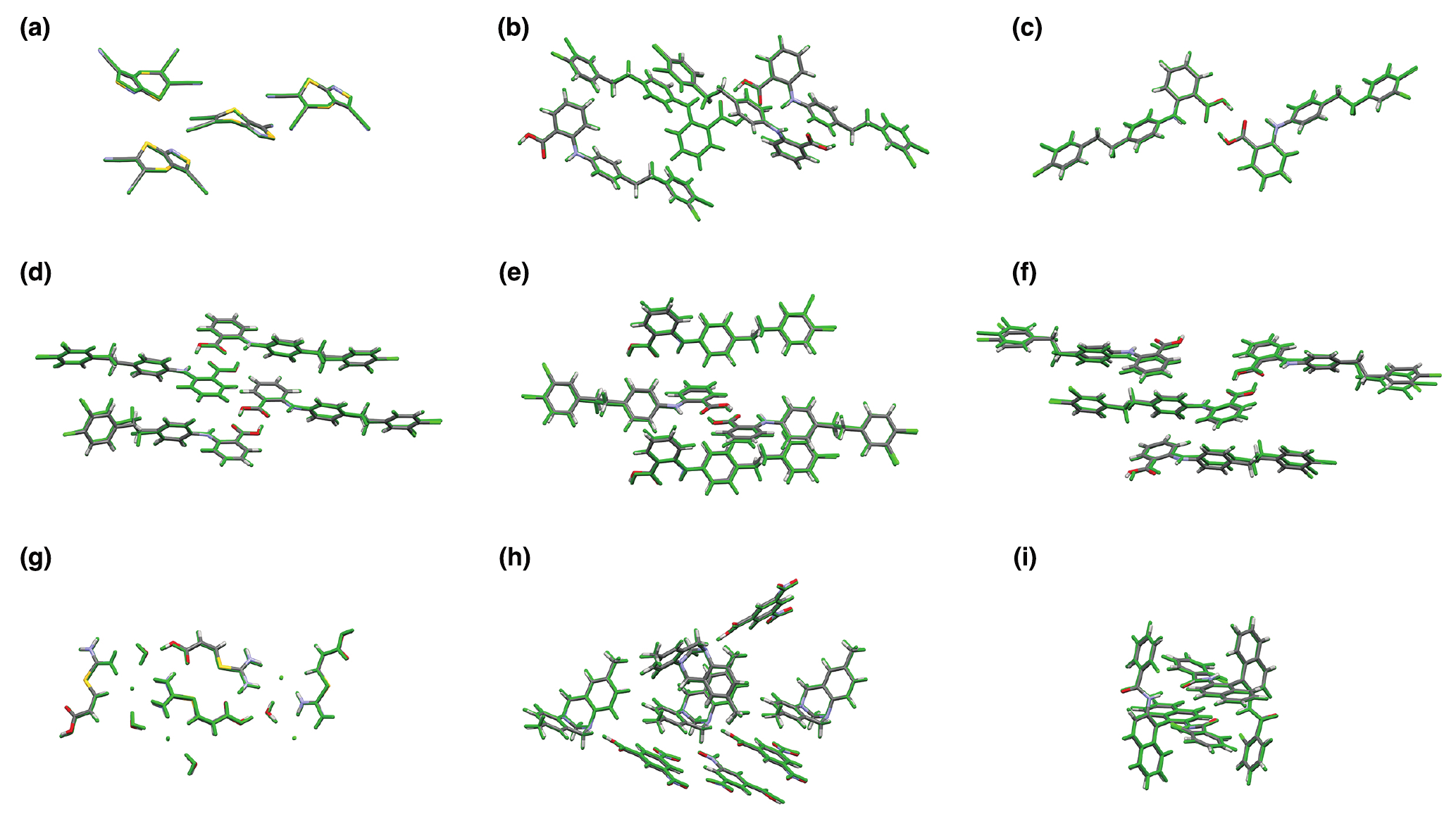}
\caption{\label{ex1}\textbf{(Extended Figure)} Overlay between the experimentally determined structures and the corresponding PBE+TS optimized structures for systems: \textbf{(a)} XXII, \textbf{(b)} XXIII-A, \textbf{(c)} XXIII-B, \textbf{(d)} XXIII-C, \textbf{(e)} XXIII-D, \textbf{(f)} XXIII-E, \textbf{(g)} XXIV, \textbf{(h)} XXV, and \textbf{(i)} XXVI.}
\end{figure*}

For all systems with only one known polymorph, the systematic and hierarchical energy ranking protocol presented herein correctly produced the experimental structure as the most stable forms (\textsc{Rank} 1). This represents a significant improvement over the \textsc{Ranks} $2$ (XXII), $2$ (XXIV), $6$ (XXV), and $1$ (XXVI) obtained by the unrefined results of Neumann \emph{et al.}, which again stresses the critical importance of an energy ranking protocol based on state-of-the-art first-principles based methodologies. For system XXIII, all experimental structures were found within the top 18 structures, with the two $Z'=2$ structures with \textsc{Ranks} $3$ (Form E) and $4$ (Form C). When only considering the $Z'=1$ structures, we find all three experimental structures among the top 10 structures, as compared to the top 26 in the initial ranking by Neumann \emph{et al.}. Moreover, all of our predicted structures agree to within $0.5$~\AA\ of the experimental structures as quantified by the root-mean-square deviation (RMSD) measure. These RMSD values are also within $0.05$~\AA\ of the initial structures obtained by Neumann \emph{et al.}, which again stresses the need for such a comprehensive crystal structure sampling protocol (see Table S4 in the SI). Overlays of the predicted and experimental structures are provided in Extended Data Fig.~\ref{ex1}.

With the exceptions of systems XXII and XXIV, our computational protocol underestimates unit cell volumes by $3.6$\% on average, which is an expected discrepancy that originates from the fact that the geometry and lattice optimizations did not include temperature (thermal expansion) effects. In this regard, a majority of the thermal expansion in molecular crystals can be accounted for \textit{via} the quasi-harmonic approximation (QHA)~\cite{Allen1969,Hoja2017,Erba2016,Heit2015}. Therefore, we have also computed the unit cell volumes at $300$~K for all of the experimental structures for XXIII (Forms A, B, C, D, E) as well as the first four $Z'=1$ structures (Str. N70, N31, N18, N42), which have yet to be experimentally observed. With this approach, we are now able to predict unit cell volumes to within $1.0$\% on average. As such, the QHA provides a simple but effective way of including thermal effects in molecular crystal structures using first-principles based methodologies. Stability rankings based on these thermally-expanded structures are shown in Fig.~\ref{fig:csp2}. The stabilities computed with the QHA can be interpreted as relative Gibbs free energies and the largest observed change stemming from the use of thermally-expanded structures amounts to $1.4$~kJ/mol at the PBE0+MBD+$F_{\rm vib}$ level.

In addition to thermal expansion, the vibrational contributions to the free energy also contain anharmonic effects, which can be accounted for by utilizing Morse oscillators (see \textit{Methods}). The corresponding free energy stability rankings with such an anharmonic treatment of the vibrational free energy are denoted by PBE0+MBD+$\widetilde{F}_{\rm vib}$ and shown in Fig.~\ref{fig:csp2}. At this level, all experimental structures are found within an energy window of only $1.5$~kJ/mol, which is well within the expected energy range for co-existing polymorphs. We note in passing that Brandenburg and Grimme have also studied the experimental structures of system XXIII utilizing a semi-empirical tight-binding approach within the QHA; however, their values lie within a much larger energy window of $\approx 8$~kJ/mol~\cite{Brandenburg2016}.

Quite interestingly, the unobserved polymorph of XXIII (Str. N70) is significantly more stable than any of the experimentally determined crystal structures, even after accounting for thermal expansion in the underlying crystal structures as well as anharmonic vibrational free energy contributions. In this regard, this polymorph is actually further stabilized by vibrational entropy and shares many structural features with form A. The most notable difference is the stacking pattern of the molecular sheets (see Figure S1). As such, we hypothesize that Form A might be kinetically favored over Str. N70 and this hitherto unobserved polymorph could potentially be crystallized by slowly melting Form A or introducing surfactants during the crystallization procedure (see SI for a detailed discussion). In addition, from a thermodynamic standpoint, Str. N18, N31, and N42 might also be observed experimentally, although Str. N42 involves a twisted molecular conformation which might not be easily accessible in solution. Experimental evidence~\cite{Reilly2016} suggests that Form A should be the most stable structure at low temperatures and Form D the most stable structure at room temperature. Indeed, we observe that Form D is stabilized by thermal effects and predicted to be more stable than Form A at the PBE0+MBD+$\widetilde{F}_{\rm vib}$ level. In addition, inclusion of anharmonic vibrational free energies brings all of the experimentally determined structures closer together, \textit{i.e.}, all of the $Z'=1$ structures are now within $0.4$~kJ/mol. 

In a broad context of crystal polymorphism, our findings suggest that late-appearing crystal forms~\cite{Bucar2015} are ubiquitous for molecules of pharmaceutical interest, further reinforcing recent experimental and computational predictions for coumarin~\cite{Shtukenberg2017}, dalcetrapib~\cite{Neumann2015}, rotigotine~\cite{Wolff2010,Bucar2015}, and ritonavir~\cite{Bauer2001}.
In the case of system XXIII, the stability of a new potential form N70 is substantially higher (by 3 kJ/mol) than that of all experimentally discovered forms. Systematic tests carried out in this work ensure the reliability of our CSP procedure to 1-2 kJ/mol, suggesting that the so far unobserved Str. N70 should be the thermodynamically stable form at ambient conditions. Obviously, experimental confirmation of this fact would be desirable and our suggestions for crystallization experiments should be useful in this endeavor. We also stress that further improvements of the presented CSP procedure are desirable and possible. For example, the efficiency could be further improved by using DFT+MBD energies for constructing tailor-made force fields during the crystal generation step of the CSP. In addition, one could improve the accuracy of free energy calculations by employing more advanced dynamical approaches, by using either path-integral molecular dynamics~\cite{Rossi2016} or the vibrational self-consistent field approach~\cite{Monserrat2013,Drummond2015,Engel2015}.

In summary, we have introduced a robust and computationally feasible procedure that yields accurate and reliable predictions of the structures and stabilities of the thermodynamically relevant polymorphs associated with complex molecular crystals, including salts, co-crystals, and flexible large molecules of pharmaceutical interest. Our approach explicitly accounts for all relevant enthalpic and entropic effects, including sophisticated treatments of Pauli exchange-repulsion, many-body dispersion interactions, and vibrational free energies at finite temperatures, all of which are directly obtained from quantum-mechanical calculations. 
The approach presented herein takes us one step closer to obtaining an enhanced fundamental understanding of polymorphic energy landscapes and routinely employing computational molecular crystal structure prediction in conjunction with experimental polymorph screening. Such a joint theoretical-experimental procedure offers a comprehensive and sustainable solution to the grand challenges associated with molecular crystals polymorphs, whose very existence offers us the promise of novel and hitherto unexplored pharmaceutical agents on one hand, and quite devastating public health and economic repercussions on the other.

\section{Acknowledgements}
J.H. and A.T. acknowledge support from the Deutsche Forschungsgemeinschaft under the program DFG-SPP 1807. H-Y.K., R.A.D., and R.C. acknowledge support from the Department of Energy (DOE) under Grant Nos. DE-SC0008626. R.A.D. also acknowledges partial support from start-up funding from Cornell University. This research used computational resources provided by the Argonne Leadership Computing Facility at Argonne National Laboratory (supported by the Office of Science of the U.S. Department of Energy under Contract No. DE- AC02-06CH11357), the National Energy Research Scientific Computing Center (supported by the Office of Science of the U.S. Department of Energy under Contract No. DE-AC02-05CH11231), the Terascale Infrastructure for Groundbreaking Research in Science and Engineering (TIGRESS) High Performance Computing Center and Visual itization Laboratory at Princeton University, the Fritz Haber Institute of the Max Planck Society (\textsc{FHI-aims} and \textsc{Mercury}), and the High Performance Computing facilities of the University of Luxembourg--{\small see \url{http://hpc.uni.lu}}.

\section{Author Contributions}
J.H., M.A.N, and A.T. designed the research. M.A.N. provided the initial crystal structures. J.H., H-Y.K., and R.A.D. carried out the polymorph ranking calculations. J.H., M.A.N., R.C., R.A.D., and A.T. analyzed the data. J.H., R.A.D., and A.T. wrote the paper with contributions from all authors.

\section{Methods}
\begin{small}
For each system in the latest blind test, we utilized the top $100$ crystal structures from the submission of Neumann \emph{et al.} (which are available in the \textit{Supplementary Information} of Ref. \citenum{Reilly2016}) as initial structures for this study. For systems with two submitted lists, we used the list which also included $Z'=2$ structures, \textit{i.e.}, structures which have two molecules in the asymmetric unit. Form E of system XXIII was the only experimental structure not present in this set and was therefore added for completeness. All calculations were performed using the all-electron \textsc{FHI-aims} code~\cite{Blum2009,Marek2014,Auckenthaler2011,Havu2009}. Throughout this work, we utilized two different accuracy levels in \textsc{FHI-aims}, which are denoted as \emph{light} and \emph{tight}. For the \textit{light} level, we use the light species default setting in \textsc{FHI-aims} for all numerical atom-centered basis functions and integration grids. The number of $k$-points ($n$) in each direction was determined by the smallest integer satisfying $n \times a \geq 25$~\AA, with $a$ being the unit cell length in a given direction. For the \textit{tight} level, we use the tight species default settings in \textsc{FHI-aims} and the number of $k$-points is determined by the criterion that $n \times a \geq 30$~\AA. Many-body dispersion (MBD) interactions were evaluated at the MBD@rsSCS level with a reciprocal-space implementation that utilized the same $k$-point mesh as the DFT calculations~\cite{Tkatchenko2012,Ambrosetti2014}. Convergence criteria of $10^{-6}$~eV, $10^{-5}$~electrons/\AA$^3$, $10^{-4}$~eV/\AA, and $10^{-3}$~eV were used for the total energy, charge density, forces, and sum of eigenvalues, respectively.

First, we performed full lattice and geometry relaxations (without any symmetry constraints) using the PBE functional~\cite{Perdew1996} in conjunction with the effective-pairwise Tkatchenko-Scheffler (TS) dispersion correction~\cite{Tkatchenko2009}, ensuring that the smallest force component is less than $0.005$~eV/\AA. Duplicate structures were identified using \textsc{Mercury}~\cite{Macrae2008}. Structures were considered similar if $20$ out of $20$ molecules within the crystals matched within $25$\% in terms of distances and within $25\degree$ in terms of angles, and the corresponding root-mean-square deviation (RMSD$_{20}$) is smaller than $0.5$~\AA. Two similar structures were considered to be identical if their PBE+TS energy (\textit{light}) agreed to within $1$~kJ/mol. Only the most stable structure among identical structures was retained throughout the protocol. These optimized structures were symmeterized using \textsc{PLATON}~\cite{Spek2009} and are provided in the SI. All structures were named according to their rank in the initial ranking by Neumann \emph{et al}. In order to determine if an experimental structure was found, we used the same settings for the crystal similarity search as described above. 

Next, relative energetic stabilities were computed based on these PBE+TS optimized structures by using PBE+TS and PBE+MBD~\cite{Tkatchenko2012,Ambrosetti2014} with \textit{tight} settings. In order to ensure the convergence of the relative energies, we have created a benchmark set consisting of $8$ small structures of system XXII and $4$ small structures of system XXIV. For these structures, PBE+MBD energies were computed using \textit{really tight} settings for the integration grids and tier $3$ basis functions. When considering all possible relative energies between structures from the same system, the mean absolute deviation (MAD) for the \textit{tight} settings amounts to only $0.1$~kJ/mol with a maximal deviation of $0.3$~kJ/mol. This illustrates the fact that \textit{tight} settings provide converged relative energies. Relative stabilities of these benchmark systems are available in Table S1 of the SI.
 
Since PBE0~\cite{Adamo1999} calculations with \textit{tight} settings are not possible for all of the studied systems due to the massive computational cost and memory requirements, we approximate the PBE0+MBD energies by adding the difference between PBE0+MBD and PBE+MBD evaluated at the \textit{light} level to the PBE+MBD energies calculated at the \textit{tight} level. For the aforementioned benchmark set, this approximation has a MAD of only $0.4$~kJ/mol with a maximum deviation of $0.8$~kJ/mol, when compared to PBE0+MBD energies evaluated with \textit{tight} settings (see Table S3 in the SI). In contrast, PBE0+MBD energies at the \textit{light} level yield a MAD of $0.8$~kJ/mol with a maximum deviation of $2.6$~kJ/mol. Therefore, our approximation provides relative energies that are in very good agreement with \textit{tight} PBE0+MBD energies. PBE0+MBD energies were computed for all structures of systems XXII, XXIII, and XXIV. For the remaining systems, PBE0+MBD calculations are available for (at least) the structures located within the top $4.5$~kJ/mol of the PBE+MBD rankings.
 
Vibrational free energies ($F_{\rm vib}$) were computed at the PBE+TS level with \textit{light} settings by utilizing the \textsc{phonopy} code~\cite{phonopy} and the finite-difference method within the harmonic approximation. The final stability rankings in Fig.~\ref{fig:ranking} were always based on PBE0+MBD+$F_{\rm vib}$ energies evaluated at temperatures correspond to the experimental crystal structure measurements. For the finite-difference calculations, we used displacements of $0.005$~\AA\ and (when necessary) supercells that ensure cell lengths greater than $10$~\AA\ in every direction. Furthermore, the vibrational free energy was evaluated in reciprocal space, where the number of $q$-points ($n$) in each direction is determined by the smallest integer satisfying $n \times a \geq 50$~\AA. All structures had no imaginary frequencies at the $\Gamma$-point and the magnitude of the three acoustic modes was smaller than $0.1$~cm$^{-1}$ in most cases  and always smaller than $0.5$~cm$^{-1}$. Vibrational free energies were calculated for (at least) all structures that are located within the top $3$~kJ/mol according to the PBE0+MBD ranking. For system XXIII, vibrational free energies were calculated for all $Z'=1$ structures and for all $Z'=2$ structures containing up to $8$ molecules per unit cell within the top $4.8$~kJ/mol of the PBE0+MBD ranking. One possibility to further improve this procedure is to include MBD interactions in the geometry optimizations and vibrational free energy calculations in addition to the static lattice energy rankings~\cite{BloodForsythe2016}. This would increase the computational cost by approximately 60\% on average and will be discussed elsewhere.

For the quasi-harmonic approximation (QHA), we performed PBE+TS lattice and geometry optimizations of several structures from system XXIII using \textit{light} settings with external hydrostatic pressures of $0.4$, $0.2$, $-0.2$, $-0.4$, and $-0.6$~GPa. Then, harmonic vibrational free energies were computed for all of the obtained structures. Based on the \textit{light} PBE+TS energies and harmonic vibrational free energies, the unit cell volume corresponding to $300$~K was determined \textit{via} the Murnaghan equation of state~\cite{Murnaghan1944}. Based on these thermally-expanded structures, the stability rankings were calculated as described above.

For all thermally-expanded structures of system XXIII, we computed the anharmonic vibrational contributions to the free energies by replacing the harmonic oscillators by Morse oscillators. This is done for all phonon modes at the $\Gamma$-point of cells containing $4$ molecules, \textit{i.e.}, for Forms A, C, D, E, and Str. N70, this corresponds to the unit cell, while for Form B and Str. N18, N31, and N42, this corresponds to a $2 \times 1 \times 1$ supercell. The structures were displaced along all normal modes in both directions, corresponding to energy changes of $0.5\,k_BT$ and $k_BT$ according to the harmonic approximation, where $k_B$ is the Boltzmann constant and $T=300$~K. The energies of all displaced structures were calculated with PBE+TS using \textit{light} settings. To have a consistent sampling of the thermally-accessible energy window, we demanded that the largest observed energy change with respect to the optimized thermally-expanded structure always lies between $k_BT$ and $1.5\,k_BT$. Therefore, the displacement amplitudes of a few low-frequency modes had to be reduced in order to sample the desired energy window. Next, we fitted a Morse potential~\cite{Morse1929,Dahl1988}, given by
\begin{equation}
V(x) = D \left(1 - e^{-a(x - x_0)} \right) ^2 , 
\end{equation}
to the obtained data points for each mode. In this expression, $x$ is the displacement amplitude, and the parameters $D$, $a$, and $x_0$ describe the well depth, the width of the potential, and the minimum of the potential, respectively. The energy of a vibrational mode in state $\nu$ can be calculates by 
\begin{equation}
E(\nu) = \hbar \omega_0 \left( \nu + \frac{1}{2} \right) - \frac{\hbar^2 \omega_0^2}{4D} \left( \nu + \frac{1}{2} \right)^2, 
\end{equation}
with 
\begin{equation}
\omega_0 = \sqrt{\frac{2a^2D}{\mu}}, 
\end{equation}
where $\mu$ is the reduced mass.
The anharmonic vibrational free energy ($\widetilde{F}_{\rm vib}$) at the $\Gamma$-point was computed according to
\begin{equation}
\widetilde{F}_{\rm vib, \Gamma} = -k_BT \ln Q_{\rm vib} , 
\end{equation}
with 
\begin{equation}
Q_{\rm vib} = \prod_i \sum_\nu \exp{\left( \frac{-E_{i,\nu}}{k_BT} \right)}, 
\end{equation}
where $i$ runs over phonon modes. This approach yields anharmonic vibrational free energies at the $\Gamma$-point for cells including $4$ molecules. In order to account also for other $q$-points, we rely on the harmonic approximation and calculate the total vibrational free energies according to:
\begin{equation}
\widetilde{F}_{\rm vib} = F_{\rm vib, full} + \widetilde{F}_{\rm vib, \Gamma} - F_{\rm vib, \Gamma} , 
\end{equation}
where $F_{\rm vib, full}$ is the fully converged harmonic vibrational free energy and $F_{\rm vib, \Gamma}$ is the harmonic vibrational free energy evaluated at the $\Gamma$ point only for the cells described above.
\end{small}

\end{document}